\documentclass[floatfix,twocolumn,preprintnumbers,aps,pra,showpacs,amsmath,amssymb,superscriptaddress]{revtex4-1}

\usepackage{hyperref} 

\usepackage{graphicx}
\usepackage{epstopdf}
\usepackage{color}
\usepackage{bm}
\usepackage{amsmath}
\usepackage{mathtools}
\usepackage{adjustbox}
\usepackage{caption}
\usepackage{subcaption}

\usepackage{epstopdf}
\newcommand{\RN}[1]{%
  \textup{\uppercase\expandafter{\romannumeral#1}}%
}

\usepackage{titlesec}
\begin{document}


\title{Quantum-Limited Squeezed Light Detection with a Camera}

\author{Elisha S. Matekole} 
\affiliation{%
Hearne Institute for Theoretical Physics, and Department of Physics and Astronomy, Louisiana State University, Baton Rouge, Louisiana 70803, USA.
}
\author{Savannah L. Cuozzo}
\affiliation{%
Department of Physics, William and Mary, Williamsburg, VA 23187, USA}
\author{Nikunjkumar Prajapati}
\affiliation{%
Department of Physics, William and Mary, Williamsburg, VA 23187, USA}
\author{Narayan Bhusal} 
\affiliation{%
Hearne Institute for Theoretical Physics, and Department of Physics and Astronomy, Louisiana State University, Baton Rouge, Louisiana 70803, USA.
}
\author{Hwang Lee} 
\affiliation{%
Hearne Institute for Theoretical Physics, and Department of Physics and Astronomy, Louisiana State University, Baton Rouge, Louisiana 70803, USA.
}
\author{Irina Novikova}
\affiliation{%
Department of Physics, William and Mary, Williamsburg, VA 23187, USA}
\author{Eugeniy E. Mikhailov}
\affiliation{%
Department of Physics, William and Mary, Williamsburg, VA 23187, USA}

\author{Jonathan P. Dowling}
\affiliation{%
Hearne Institute for Theoretical Physics, and Department of Physics and Astronomy, Louisiana State University, Baton Rouge, Louisiana 70803, USA.
}
\affiliation{%
	NYU-ECNU Institute of Physics at NYU Shanghai, 3663 Zhongshan Road North, Shanghai, 200062, China.
	}
\affiliation{%
CAS-Alibaba Quantum Computing Laboratory, CAS Center for Excellence in Quantum Information and Quantum Physics, University of Science and Technology of China, Shanghai 201315, China.
    }
\affiliation{%
National Institute of Information and Communications Technology,
4-2-1, Nukui-Kitamachi, Koganei, Tokyo 184-8795, Japan
    }

\author{Lior Cohen} 
\affiliation{%
Hearne Institute for Theoretical Physics, and Department of Physics and Astronomy, Louisiana State University, Baton Rouge, Louisiana 70803, USA.
}
\affiliation{%
cohen1@lsu.edu}

\date{\today}

\begin{abstract}
We present a technique for squeezed light detection based on direct imaging of the displaced-squeezed-vacuum state using a CCD camera. We show that the squeezing parameter can be accurately estimated using only the first two moments of the recorded pixel-to-pixel photon fluctuation statistics, with accuracy that rivals that of the standard squeezing detection methods such as a balanced homodyne detection. Finally, we numerically simulate the camera operation, reproducing the noisy experimental results with low signal samplings and confirming the theory with high signal samplings.
\end{abstract}


\maketitle

\textit{Introduction}.--- Squeezed light is an optical state in which the fluctuations of one quadrature are suppressed below the shot noise limit (SNL)~\cite{GK2005,  Walls1, TRalph, Mlynek97, Fabre92, Kim92, Leo97, Xiao87,Lvovsky}.
It has become an important resource in the field of quantum optics and quantum information, as more and more optical technologies are crossing the boundary into the quantum realm. Squeezed states have been successfully applied in continuous-variable quantum communication protocols \cite{polzik_book,lvovsky09,andersenLPR2010} and in improving performance of optical sensors~\cite{quant_sense:pooser:2018}, including gravitational wave detectors~\cite{GW_detect_with_squeezed:Chau:2014}. Numerous methods for generation of squeezed light have developed based on variety of nonlinear materials~\cite{TRalph,Lvovsky}. The common ones utilize parametric down conversion in nonlinear crystals~\cite{GK2005,Walls1,cohen2018absolute} although atom-based sources based on a polarization self-rotation effect \cite{Matsko2002,Mikhail2008,Barre2011,Agha2010,Ries2003} and four-wave mixing \cite{Walls1985, slush1985, kolo1994, lett2007,Boyer2015} are also being pursued. 

The detection of squeezed light is usually carried out in one of the three ways: by direct intensity detection or photon counting (for intensity-squeezed light only), using a phase-shifting cavity~\cite{TRalph}, and by far the most common among the three, homodyne or heterodyne detection by beating the squeezed light field with a classical local oscillator. In this letter, we present a technique that allows us to characterize the squeezing parameter in a displaced squeezed vacuum state employing a CCD camera without using correlation detection. We demonstrate that the amount of squeezing can be derived from the first and second moments of the photon statistics per pixel, with the accuracy similar to what would be achieved with homodyne detection. At the same time, the proposed method may be particularly beneficial in applications of squeezing to enhance optical imaging ~\cite{quantum_imaging1,Genovese_2016}.

\begin{figure}[tb]
\centering
   \includegraphics[trim=20 0 0 0 ,angle=0 ,width=1.0\columnwidth]{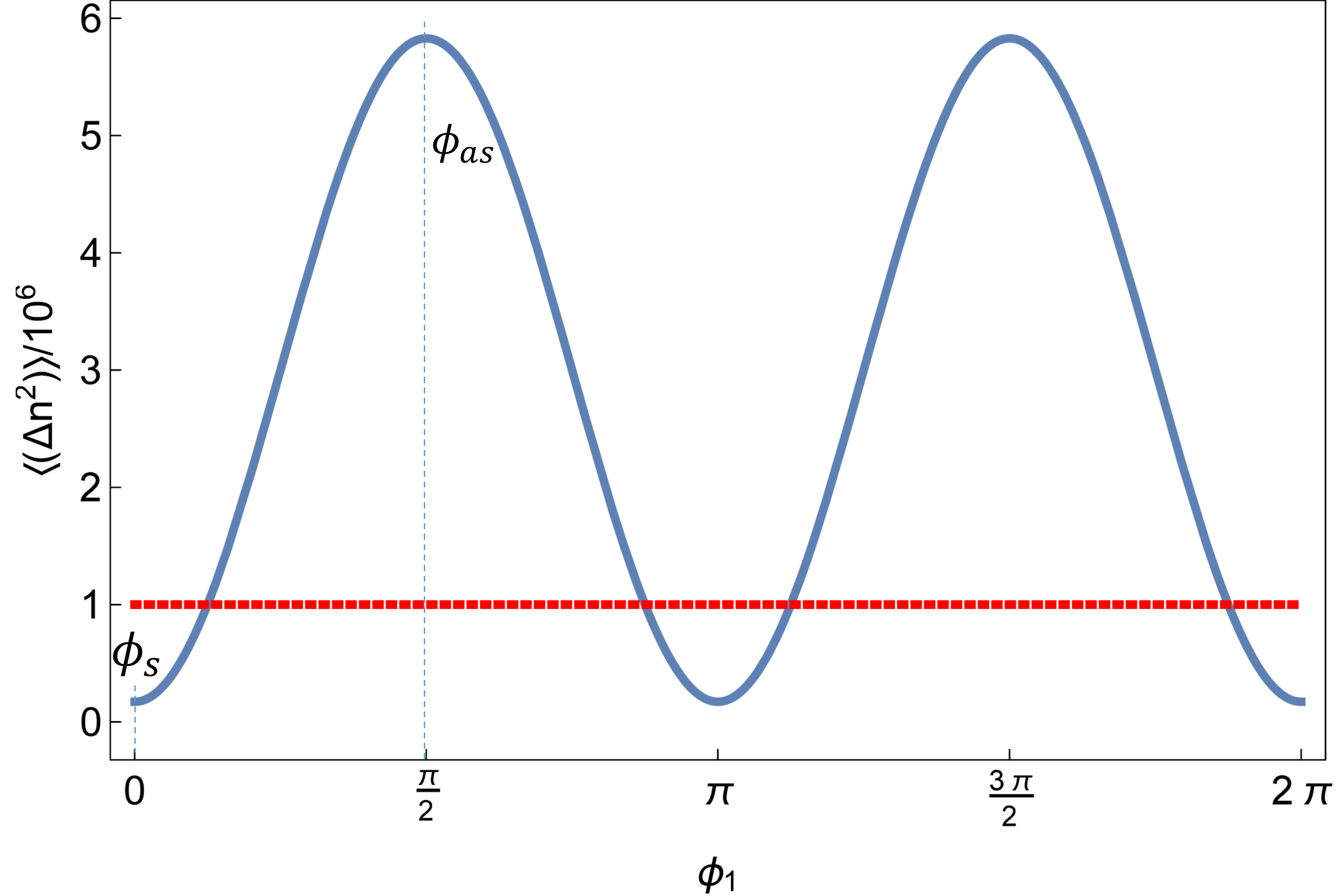}
   
\caption{(Color Online)Phase dependent displaced squeezed light.  Plot of photon number fluctuation as a function of $\phi_{1}$. The constant line represents the shot-noise limit. When the photon-number fluctuation is below the SNL the field is squeezed. The average number of photons in the pump beam is $\bar{n}_{\textrm{pump}}=10^{10}$, $\theta=10^{-2}$, the leaked amount of pump photons is $\bar{n}_{\alpha}=|\alpha\theta|^{2}$, the average number of photon in squeezed vacuum generated is $\bar{n}_{s}=1$. The values of $\phi_{1}$ represented by $\phi_{\rm{as}}$, and $\phi_{\rm{s}}$ represent anti-squeezing, and squeezing respectively.}
\label{setup}
\end{figure}

\begin{figure}[t!]
\centering
\begin{subfigure}[b]{1\columnwidth}\includegraphics[angle=0,width=0.8\columnwidth]{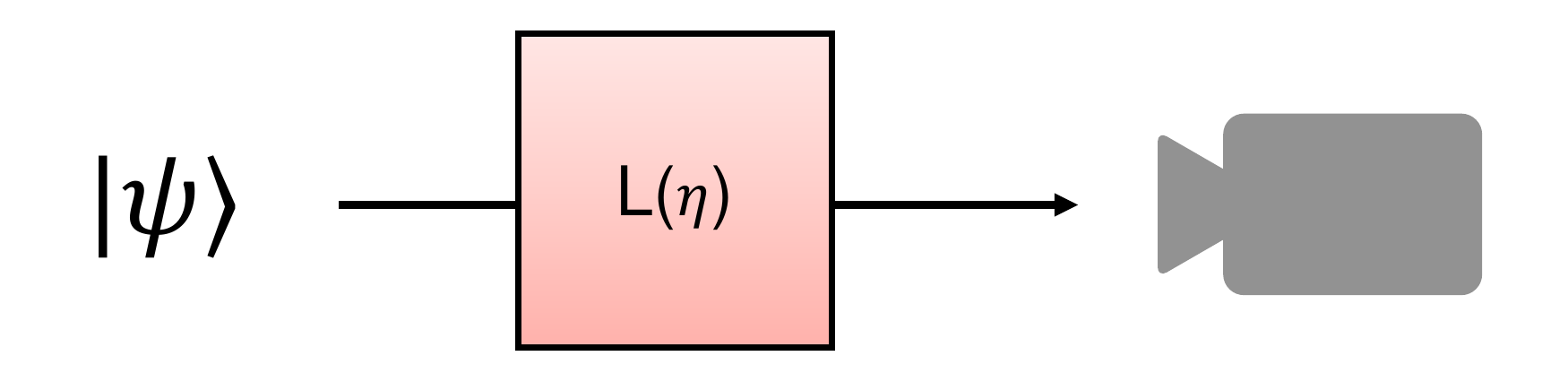}
   \caption{}
   \label{fig:2a}
\end{subfigure}
\begin{subfigure}[b]{1\columnwidth}
   \includegraphics[trim=0 10 0 0,clip,angle=0,width=1\columnwidth]{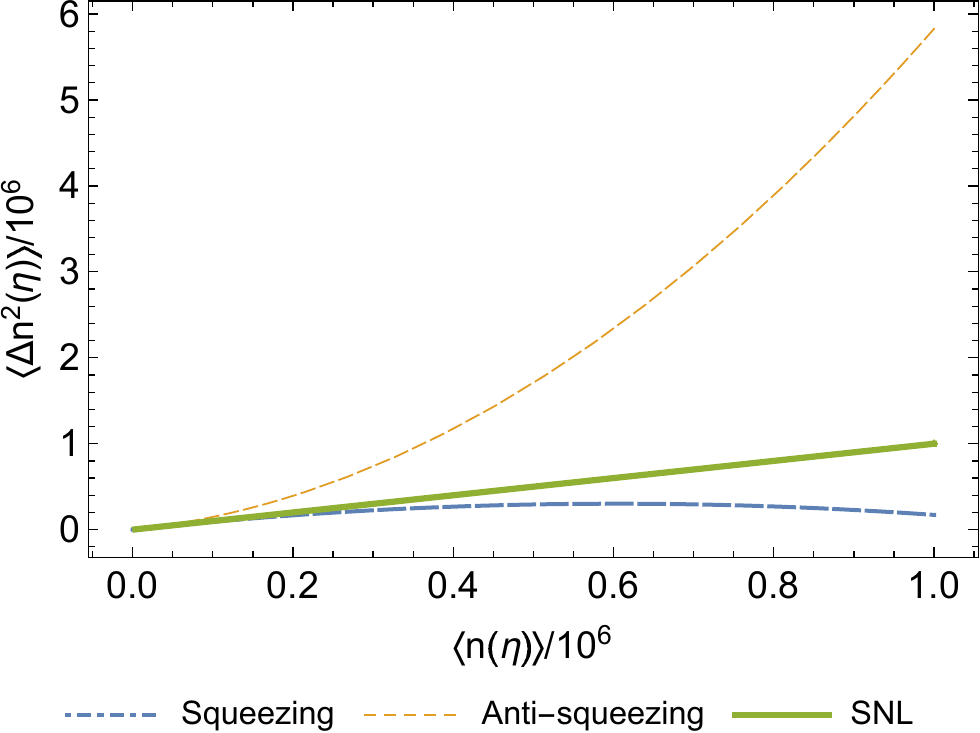}
   \caption{}
   \label{fig:2b}
\end{subfigure}
\caption{(Color Online) Theoretical model and results (a) Building blocks of the model as a two mode propagator. The squeezed light $|\xi\rangle$, is mixed with a strong coherent light $|\alpha e^{i\phi_{1}}\rangle$, on an unbalanced beamsplitter with reflectivity of $\theta<<1$. After tracing the pump mode, the final output state can be approximated as $|\alpha e ^{i\phi_{1}}\theta,\xi\rangle$. The statistical properties of a single pixel of the camera are simulated by an adjustable attenuation, ${\rm L}(\eta)$ (b) Parametric plot of photon-number fluctuation and photon counts, of parameter $\eta$, the transmission. The dotted curve represents the squeezing as the variance in photon counts is less than the average photon counts, ($\langle\Delta\hat{n}^{2}\rangle < \langle \hat{n} \rangle$). The dashed curve shows anti-squeezing since the variance is greater than the average photon counts, ($\langle\Delta\hat{n}^{2}\rangle > \langle \hat{n} \rangle$). The solid curve represents the shot-noise limit obtained from coherent light, ($\langle\Delta\hat{n}^{2}\rangle = \langle \hat{n} \rangle$).}
\label{fig2}
\end{figure}

\textit{Method}.---  We mix the strong pump with squeezed vacuum light $|\xi\rangle$ at an unbalanced beam splitter of reflectivity $\theta<<1$ for the pump field. The pump is a coherent light state $|\alpha e^{i \phi_{1}}\rangle$ with $\bar{n}_{\textrm{pump}}=|\alpha|^{2}$ as the average number of photons. 
The phase $\phi_{1}$ is the controllable phase shift between $|\alpha\rangle$ and $|\xi\rangle$, which takes the state of the resulting field from squeezing to anti-squeezing. Though we present the displacement here as a separate and active operation, in reality it commonly comes for free \cite{aggarwal2018room}. In other setups, the pump co-propagates with the squeezed light \cite{Mikhail2008, cohen2018absolute} and thus, the squeezed light is displaced by default.

The output states containing mostly squeezing after the beam splitter can be approximated to $|\alpha e^{i \phi_{1}} \sin\theta\cos\theta,\xi\cos^{2}\theta\rangle$ and $|\alpha e^{i \phi_{1}}\cos^{2}\theta,\xi\cos\theta\sin\theta\rangle$,
where $|\alpha,\xi\rangle=\hat{D} (\alpha) \hat{S} (\xi)|0\rangle$, $\hat{D}$ is the displacement operator, $\hat{S}$ is the squeezing operator. The exact state is entangled and the approximation is valid for $\theta<<1$. 
The final output state is a displaced squeezed vacuum state, approximated as $|\psi\rangle=|\alpha\theta e^{i \phi_{1}},\xi\rangle$.
Next, we show the dependence of this final output state on the phase, $\phi_{1}$ at which squeezing and anti-squeezing occurs. We calculate the amplitude fluctuation, $\langle\Delta \hat{n}\rangle^2$ as a function of the phase $\phi_{1}$ as,

\begin{align}
\label{delnp1}
&\langle \Delta \hat{n}^{2}\rangle=\bar{n}_{\alpha}+2\bar{n}_{\alpha}\bar{n}_{s}+2\bar{n}_{s}+2\bar{n}_{s}^{2}\nonumber\\
& -2 \cos{(2\phi_{1})}\bar{n}_{\alpha}\sqrt{\bar{n}_{s}(1+\bar{n}_{s})},
\end{align}
which shows that the sub-shot noise suppression happens only in the presence of squeezed light. Here $\bar{n}_{\alpha}$ represents average leaked pump photons, and $\bar{n}_{s}$, the average number photons in squeezed vacuum state.

 
In Fig.~\ref{setup}, we show the fluctuation in photon counts $\langle \Delta \hat{n}^2\rangle$ as a function of $\phi_{1}$ for the output squeezed state and compare it against the shot-noise limit $\langle \Delta \hat{n}^2\rangle={\bar{n}_{\alpha}+\bar{n}_{s}}$, where $\bar{n}_{s}=\sinh^{2}{r}$, and $r$ is the squeezing parameter. We observe squeezing and anti-squeezing for $\phi_{1}=\phi_{\rm{s}}=0$, and $\phi_{1}=\phi_{\rm{as}}=\pi/2$ respectively. This confirms that even though almost all the photons are from the coherent light, the overall non-classical statistics are still preserved. 
\begin{figure}[t]
\centering
\begin{subfigure}[b]{0.53\columnwidth}
   \includegraphics[angle=0,width=1\columnwidth]{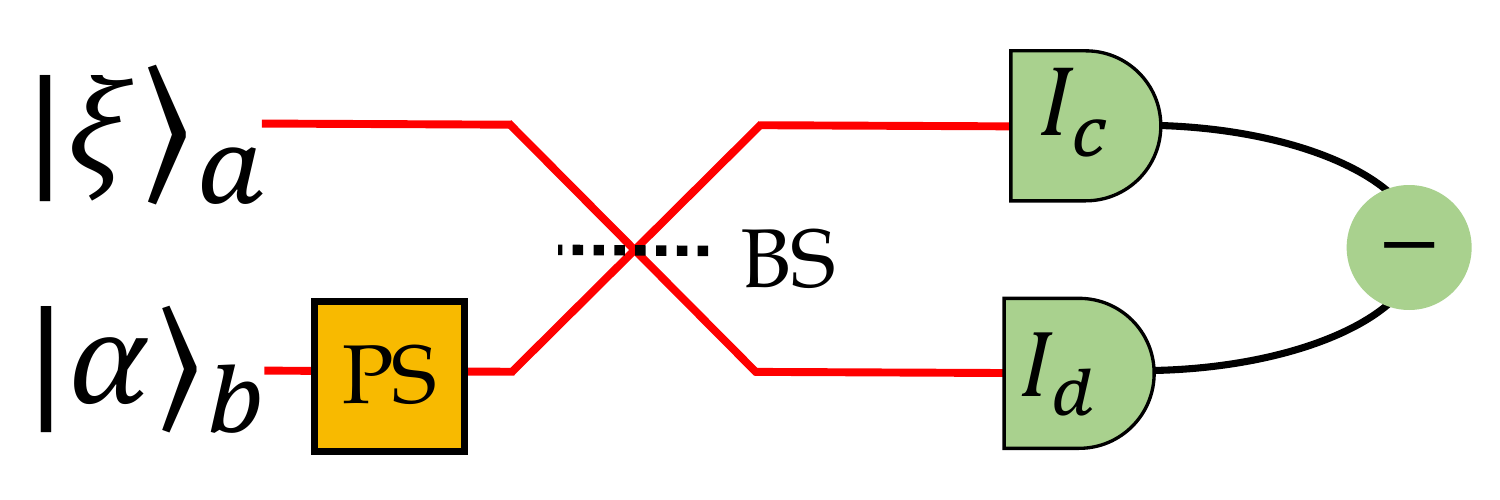}
   \caption{}
   \label{homodyne}
\end{subfigure}
\begin{subfigure}[b]{0.45\columnwidth}
   \includegraphics[trim=0 -0 0 0,angle=0,width=1\columnwidth]{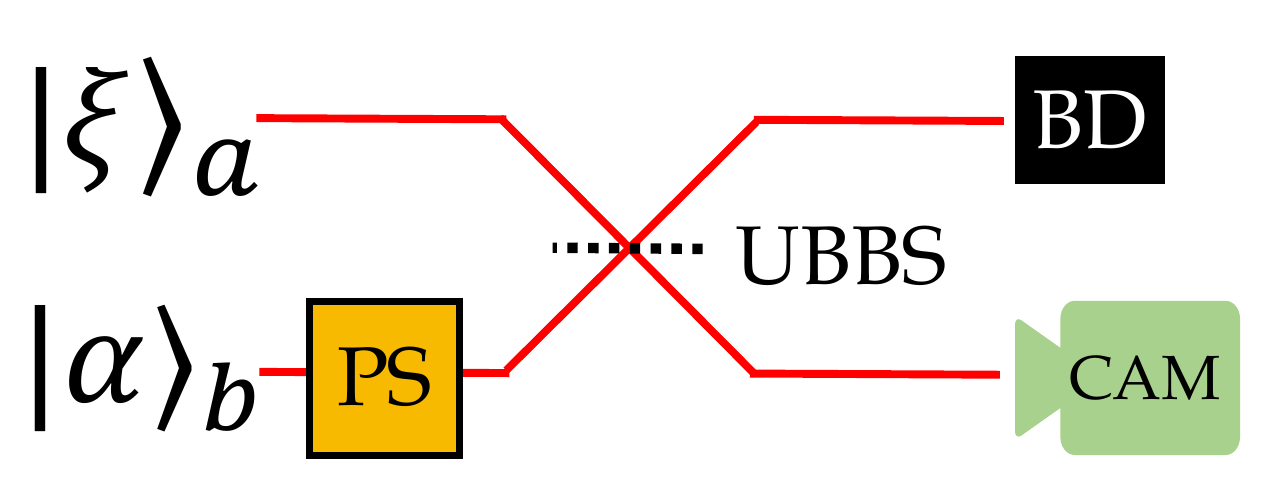}
   \caption{}
   \label{fig:3b}
\end{subfigure}
\caption{(Color Online) Schematic diagrams of squeezed-light detector (a) Balanced homodyne detection, the common method of squeezed-light detection. Here the local oscillator (L.O.) enters along $\hat{b}$, and $\bar{n}_{\rm L.O.}=10^{6}$. The squeezed light field enters along $\hat{a}$, and $\bar{n}_{s}=1$. (b) Our proposed setup comprises field displacement and a single-photon camera. PS - phase shifter, BS - beamsplitter, UBBS - unbalance beamsplitter, BD - beam dumper and CAM - camera. }
\label{homodyne}
\end{figure}


In Fig.~\ref{fig:2a}, we present the analytical model of our method to detect squeezed light using a camera. The mixing of the squeezed and coherent light is modeled by beam-splitter transformation with reflectivity $\theta$. We consider $\theta\sim 10^{-2}$, so that we lose only a small amount of squeezed light.
After the beam splitter, the light is incident on the camera. Since we focus on a single pixel, the camera acts as a tunable attenuator, L($\eta$) of transmission $\eta$. This conjecture is proven in the supplementary information. However intuitively, when looking on one pixel, the rest of the light, incident on the other pixels, is lost. 
We look at the statistics of the detected signal photons per pixel and plot the first two moments of the photon counts against each other, for $\phi_{\rm{s}}=0$, and $\phi_{\rm{as}}=\pi/2$ where maximum squeezing, and maximum anti-squeezing happens as noted from Fig.~\ref{setup}. 


For coherent light state we know that $\langle\Delta \hat{n}^{2}\rangle=\langle \hat{n} \rangle$. Therefore, the condition for squeezing is defined as $\langle\Delta \hat{n}^{2}\rangle < \langle \hat{n} \rangle$, and anti-squeezing $\langle\Delta \hat{n}^{2}\rangle > \langle \hat{n} \rangle$  as seen in the Fig.~\ref{fig:2b}. Since we have the first two moments for the displaced squeezed light state, the stage is now set to extract the squeezing parameter. 

\textit{Analytical Results}.--- Next, we demonstrate the extraction of squeezing, and the amount of coherent light from Fig.~\ref{fig:2b}. 
The analytical expressions for the average photon number, and the variance per pixel  is shown in Eq.~\ref{counts}.
We plot the variance versus the average photon counts as a function of the transmission, $\eta$.
\begin{align}
\label{counts}
\langle \hat{n} \rangle&=\eta(\bar{n}_{\alpha}+\bar{n}_{s})\nonumber\\
\langle\Delta \hat{n}^2\rangle&=\frac{1}{2}\eta\Big(2\bar{n}_{\alpha}(1+\bar{n}_{s})+\bar{n}_{s}(2+\bar{n}_{s})\nonumber\\
&-4\eta\bar{n}_{\alpha}\sqrt{\bar{n}_{s}(\bar{n}_{s}+1)}\cos{2\phi_{1}}\nonumber\\
&+\bar{n}_{s}(1+2\bar{n}_{\alpha}+2\bar{n}_{s})(2\eta-1)\Big).
\end{align}

We find the curve fit for $\phi_{1}=\phi_{\rm{s}}=0$, and $\phi_{1}=\phi_{\rm{as}}=\frac{\pi}{2}$, which are the values for squeezing and anti-squeezing respectively from Fig.~\ref{fig:2b}, and extract the values of $\bar{n}_{s}$ and $\bar{n}_{\alpha}$.   
The curve fit obeys the equation $\langle\Delta \hat{n}^{2}(\eta)\rangle=\langle \hat{n}(\eta)\rangle+q\langle \hat{n}(\eta)\rangle^{2}$. Since we have two unknowns we need two equations. Therefore to be able to extract $\bar{n}_{s}$, and $\bar{n}_{\alpha}$ we use the coefficient $q$ for both the squeezing and anti-squeezing curves, where q is given by;
\begin{align}
\label{curveparamt}
q =&\frac{1}{(\bar{n}_{\alpha}+\bar{n}_{s})^{2}}\Big(\bar{n}_{s}(1+2\bar{n}_{\alpha}+2\bar{n}_{s})\nonumber\\
&-2\bar{n}_{\alpha}\sqrt{\bar{n}_{s}(1+\bar{n}_{s})}\cos2\phi_{1}\Big).
\end{align}
$q<0$ is a sign of sub-Poissonian photon statistics and thus $q$ is a measure of quantum effect \cite{fox2006quantum}. 

Next we compare the sensitivity of obtaining $\bar{n}_{s}$ using our method with the homodyne method. 
In Fig.~\ref{homodyne}, we sketch the homodyne and our setup. In our homodyne scheme the signal is the average variance in the field quadrature, $\langle\Delta \hat{X}(\phi_{1})^{2}\rangle$, and the noise is given by the variance of the signal, $\langle\Delta(\Delta \hat{X}(\phi_{1})^{2})^{2} \rangle$, which is the variance of the variance in the field quadrature.  For Gaussian probability statistics there is a connection between the second and fourth moment, where the latter is twice the square of the former. This fact is also useful experimentally where it is hard to measure the fourth moment. Therefore the sensitivity in the value of $\bar{n}_{s}$ can be extracted from homodyne detection as follows,
\begin{align}
    {\label{homo_ns}}
    &\langle(\Delta\hat{X}(\phi_{1}))^{2}\rangle=\frac{1}{2}(2\bar{n}_{s}+1-2\sqrt{\bar{n}_{s}(\bar{n}_{s}+1)}\cos{\phi_{1}}),\\
     &\langle(\Delta\bar{n}_{s})^{2}\rangle=\frac{\langle(\Delta\hat{X}(\phi_{1}))^{2}\rangle^{2}}{|\frac{\delta \langle(\Delta\hat{X}(\phi_{1}))^{2}\rangle }{\delta\bar{n}_{s}}|^{2}},\\
    &\langle(\Delta\bar{n}_{s})^{2}\rangle=2\bar{n}_{s}(\bar{n}_{s}+1),
\end{align}
where the sensitivity is phase independent.

Similarly, we calculate the sensitivity of $\bar{n}_{s}$ for our method. Here, the $\bar{n}_{s}$ information is encoded in the curve-fit parameter $\textit {q}$ as shown in Eq. \ref{curveparamt}. First we rewrite $\textit {q}$ in terms of $\langle\hat{n}_{1}(\eta)\rangle$ and $\langle(\Delta\hat{n}_{1}(\eta))^{2}\rangle$ using the curve fit equation as, 
\begin{align}
    q=\frac{\langle(\Delta\hat{n}_{1}(\eta))^{2}\rangle-\langle\hat{n}_{1}(\eta)\rangle}{\langle\hat{n}_{1}(\eta)\rangle^{2}}.
\end{align}
Using the error propagation we get,
\begin{align}
    {\label{camera_ns}}
    \langle\Delta q^{2}\rangle&=(\frac{\delta q}{\delta \langle\hat{n}_{1}(\eta)\rangle})^{2}\langle(\Delta\hat{n}_{1}(\eta))^{2}\rangle\nonumber\\
    &+(\frac{\delta q}{\delta \langle(\Delta\hat{n}_{1}(\eta))^{2}\rangle})^{2}\langle\Delta(\Delta\hat{n}_{1}(\eta)^{2})^{2}\rangle,\\
    \langle\Delta\bar{n}_{s}^{2}\rangle&=\frac{\langle\Delta q^{2}\rangle}{|\frac{\delta q}{\delta \bar{n}_{s}}|^{2}}.
    \end{align}
In the limit of $\bar{n}_{\alpha}>>\bar{n}_{s},1$, the sensitivity of $\bar{n}_{s}$ is the same as homodyne method,    
    \begin{align}
    {\label{camera_ns1}}
   &\nonumber\langle\Delta\bar{n}_{s}^{2}\rangle\approx\frac{4\bar{n}_{\alpha}^{4}\bar{n}_{s}(1+\bar{n}_{s})(2\bar{n}_{s}+1\pm 2\sqrt{\bar{n}_{s}(1+\bar{n}_{s})})^{2}}{2\bar{n}_{\alpha}^{4}(2\sqrt{\bar{n}_{s}(1+\bar{n}_{s)}}\pm(2\bar{n}_{s}+1))^{2}},\\
&= 2\bar{n}_{s} (\bar{n}_{s}+1). 
\end{align}
where $\pm$ stands for the squeezed and anti-squeezed phases.

It is interesting to see that the noise in calculating $\bar{n}_{s}$ is the $\emph{quantum noise of squeezed light state}$. Hence from Eqns. \ref{homo_ns},~\ref{camera_ns1}, we conclude that the camera method performs as good as the widely used homodyne method for squeezed light detection. Also, our camera method is quantum limited. 



\begin{figure*}[tb!]
\centering
   
\begin{subfigure}[b]{0.65\columnwidth}
   \includegraphics[angle=0,width=1\columnwidth]{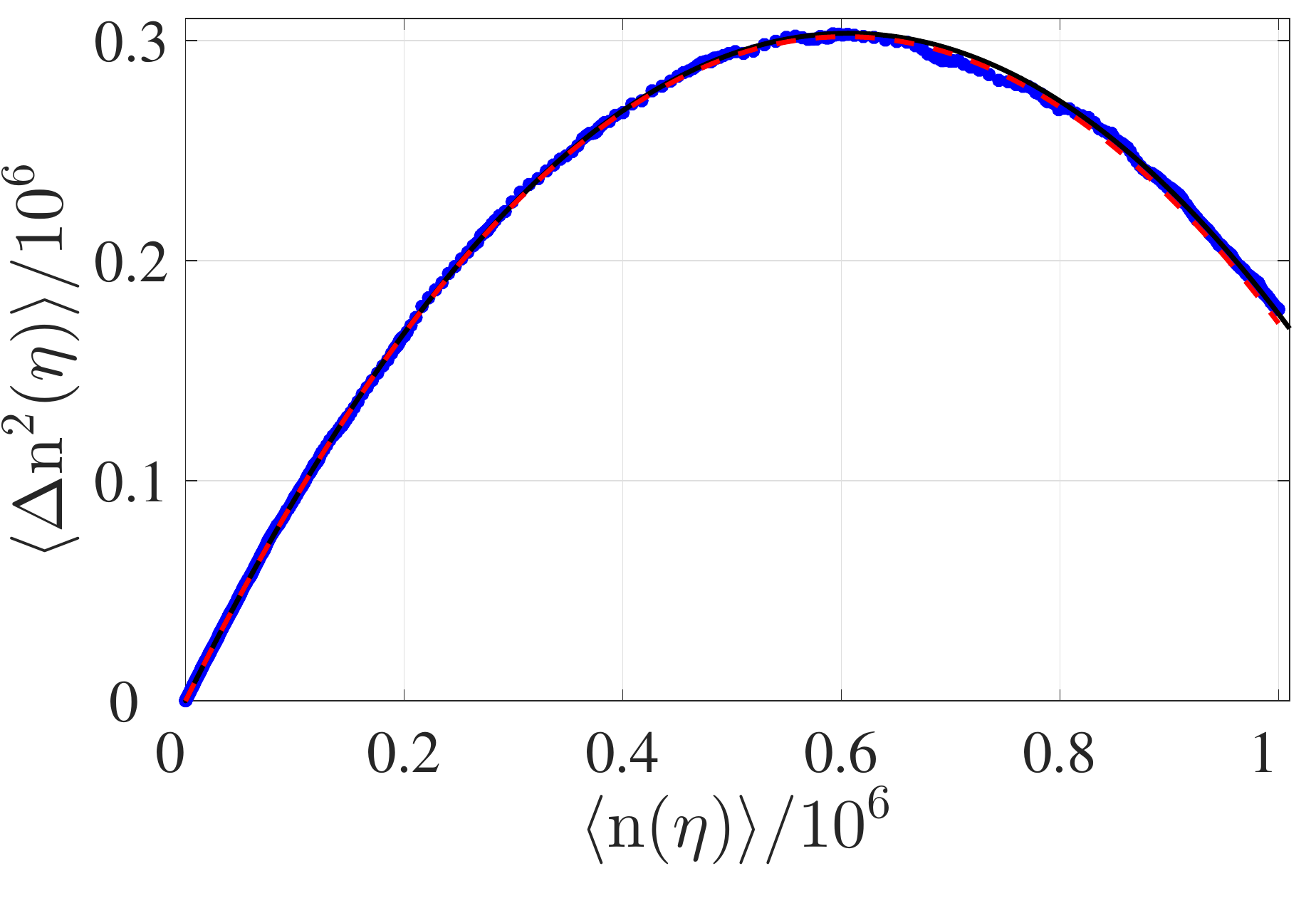}
   \caption{}
   \label{fig:4c}
\end{subfigure}
\begin{subfigure}[b]{0.63\columnwidth}
   \includegraphics[angle=0,width=1\columnwidth]{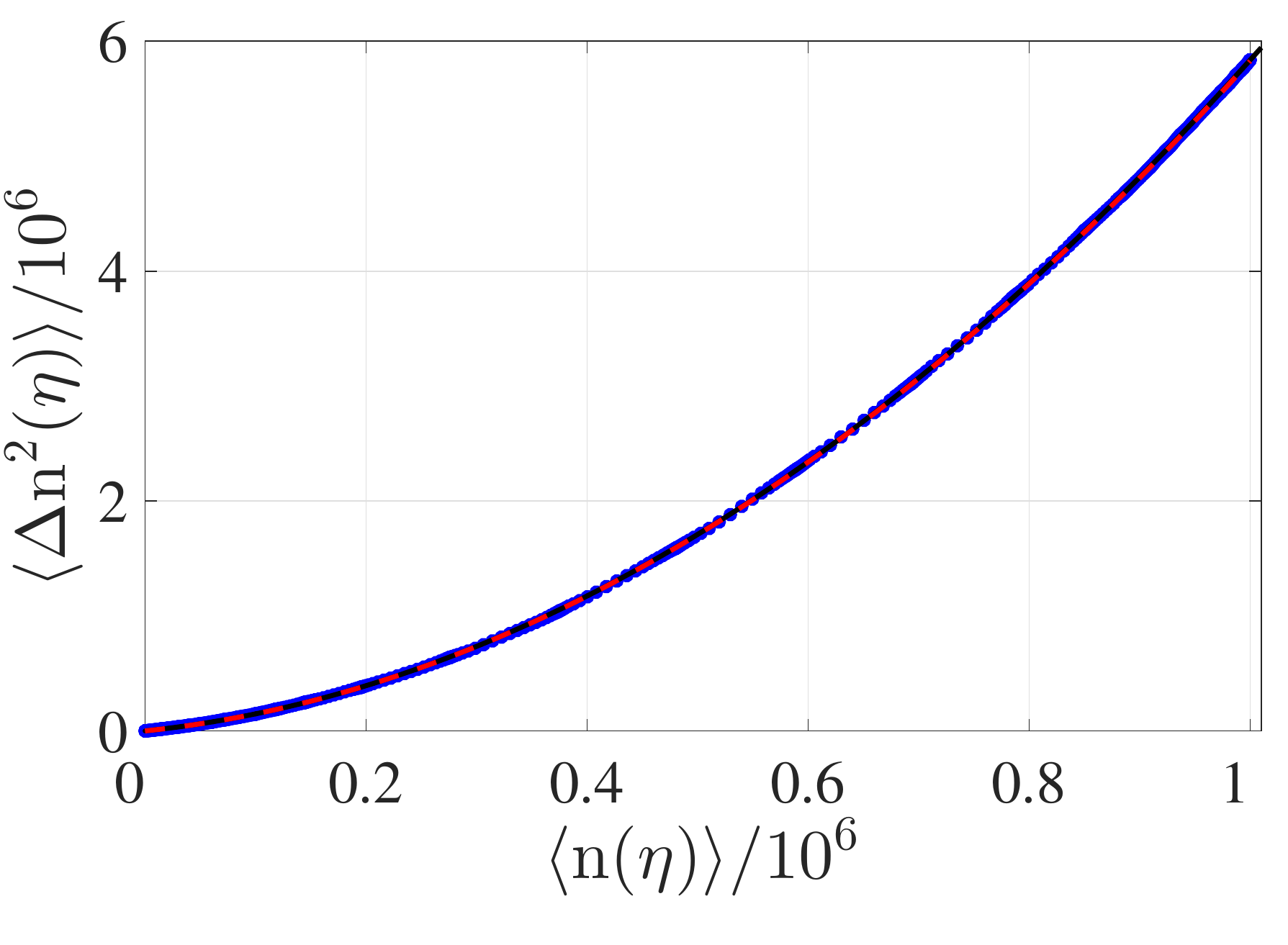}
   \caption{}
   \label{fig:4f}
\end{subfigure}
\begin{subfigure}[b]{0.67\columnwidth}
   \includegraphics[clip,trim=0 -10 0 0, angle=0, width=1\columnwidth] {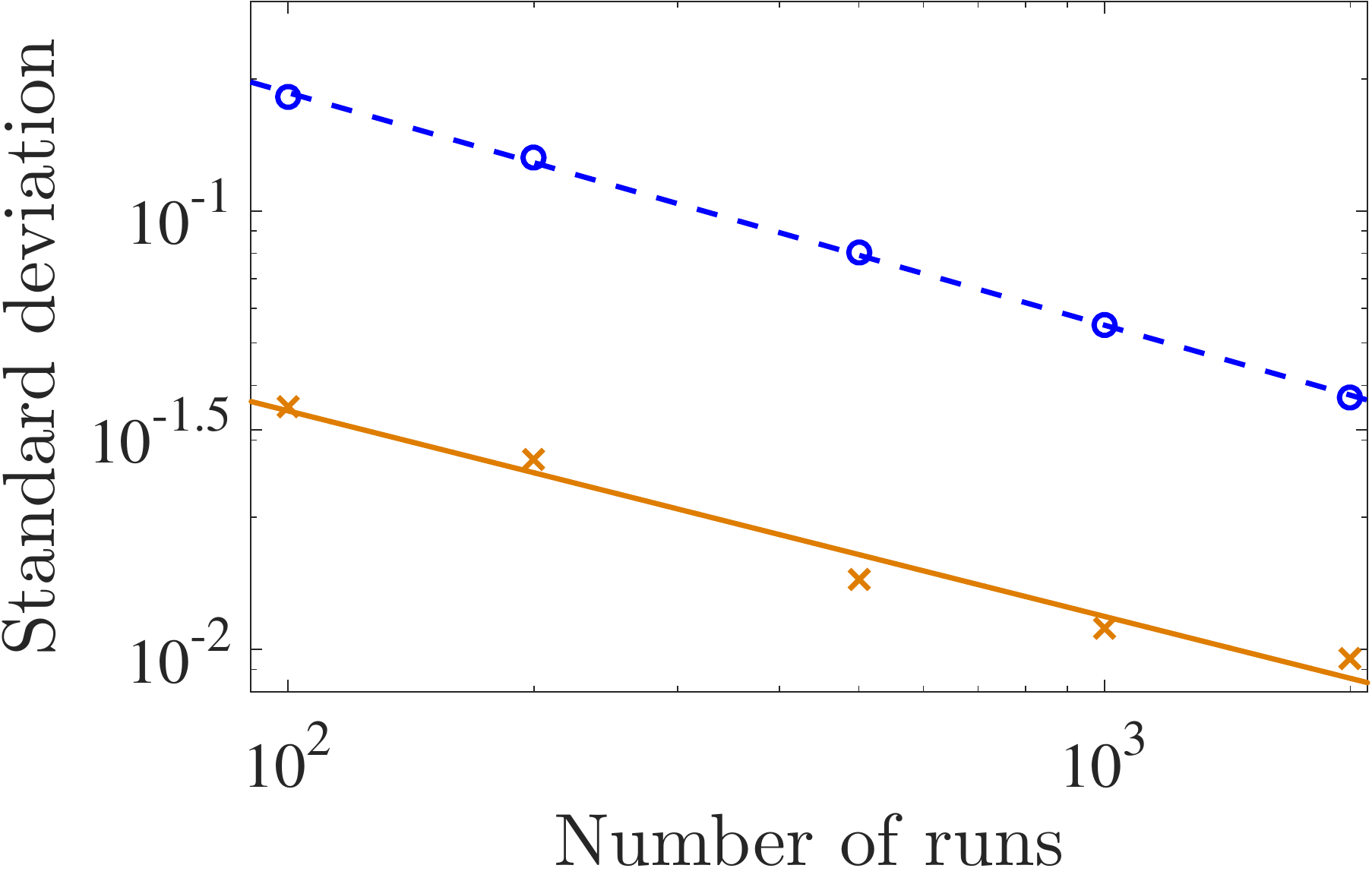}
   \caption{}
   \label{fig:4g}
\end{subfigure}
\caption{(Color Online) The simulation results of the variance as a function of the intensity for 10,000 iterations with squeezed \textbf{(a)} and \textbf{(b)} anti-squeezed states. The  results are shown after integrating over pixels. The state parameters are; $\bar n_\alpha = 10^6$ and $\bar n_s = 1$. The solid black line is a fit to second order polynomial, and the dashed red line is the theoretical relation, obtained from Eq. \ref{counts}. The standard deviation of the fit parameter $q$ is plotted as a function of number of simulation runs \textbf{(c)}, for squeezed (orange crosses) and anti-squeezed simulations. The lines are linear fit. The slopes are $-0.47 \pm 0.075$, $-0.53 \pm 0.015$  for squeezed and anti-squeezed data.} 
\label{fig4}
\end{figure*}

\textit{Simulations}.--- 
In order to demonstrate our method we simulate the experiment of measuring squeezed state with a camera. A detailed simulation procedure and mathematical framework is described in the supplemental information. In short, a photon number is randomly picked, according to the state photon statistics. The photons are distributed to 32-by-32 camera pixels. After repeating the simulation $N$ times, the intensity and variance are computed for each pixel. The variance can be plotted as a function of intensity where each point in the plot is represented by a different pixel (See Fig. S2 in the supplemental information). 

To increase the precision of the results (without adding more data), one can integrate or group pixels. It can be done in many ways and here we choose to integrate over pixels such that the first point is the first pixel, the second point sums over the first two pixels, the third on three, and so on. The last point sums over all of the pixels. By doing that, we improved the fitting error and the results are $(-8.242 \pm 0.001)\times10^{-7}$, and $(4.837 \pm 0.001)\times10^{-6}$ for squeezed (Fig. \ref{fig:4c}) and anti-squeezed (Fig. \ref{fig:4f}) states, respectively. The values are very close to the theoretical values of $-8.2842\times10^{-7}$ and $4.8284\times10^{-6} $, giving the values of $\bar{n}_{\alpha}=(1.0065\pm 0.002)\times10^{6}$, and $\bar{n}_{s}=1.0098\pm0.00045$. 
The slight deviation can be explained by the quasi-random-number generator, which probably introduces correlations in the random numbers, which in turns add observed non-statistical noise to the results. 

Figure \ref{fig:4g} shows the precision (standard deviation - SD) of the value $q$ as a function of the number of runs. The precision is improved as one over the square root of the number of runs, as expected. For anti-squeezing, the precision is about five times worse than for squeezing. It is a result of the more spread in the photon statistics which adds more noise to the simulations. Quantitatively, the SD of the anti-squeezing photon statistics is 5.8 larger than the SD of the squeezing photon statistics (see Eq. \ref{delnp1}), which coincides with the factor of five as seen in Fig. \ref{fig:4g}. 

\textit{Conclusion}.--- 
We have proposed a scheme to detect single-mode squeezed light without using the homodyne detection. We mix the squeezed light with a strong coherent light field at an unbalanced beam splitter. The final state is a displaced-squeezed vacuum state with a controllable phase shift $\phi_{1}$. 
We demonstrate that the amount of squeezing can be estimated from the first two moments of the photon statistics obtained from the camera. 
We show that our method does equally well compared to homodyne detection, and is quantum-limited.
Lastly, we carry out a numerical simulation of our model to calculate the amount of squeezing, and compare with the analytical results. We find that our numerical simulation results agree with the analytical results.

\textit{Aknowledgements}.---
This work was completed shortly after the demise of one of the authors, Jonathan P. Dowling. It is with immense gratitude that we acknowledge his guidance, and vision. His approach to research, not to mention his humor and wit will be sorely missed. 
This research was supported by the grant AFOSR FA9550-19-1-0066.  E.S.M., L.C., N.B., H.L. and J.P.D. would like to acknowledge the Air Force Office of Scientific Research, the Army Research Office, the Defense Advanced Research Projects Agency, and the National Science Foundation.  S.L.C., N.P., E.E.M and I.N. would like to acknowledge the support of AFOSR DURIP FA9550-16-1-0417 grant. S.L.C. and N.P. acknowledge the support of the Virginia Space Grant Consortium (VSGC) Graduate Research STEM Fellowship Program.
\bibliography{sqzd}
\end{document}